\documentclass[aps,prd,reprint,preprintnumbers,floats,epsfig,nofootinbib,amssymb,
nofootinbib]{revtex4}

\usepackage{slashed}
\usepackage{graphicx,color}
\usepackage{epsfig}
\usepackage{subfigure}
\usepackage{epsfig}
\usepackage{dcolumn}
\usepackage{bm}
\usepackage{color}
\usepackage{ulem}
\usepackage{enumitem}
\usepackage[latin1]{inputenc}
\usepackage[colorlinks,
            linkcolor=black,
            anchorcolor=black,
            citecolor=black
            ]{hyperref}

\begin{document}

\baselineskip=15pt


\title{DFSZ Axion Couplings Revisited}

\author{Jin Sun${}^{1}$\footnote{019072910096@sjtu.edu.cn}}
\author{Xiao-Gang He${}^{2,3}$\footnote{hexg@phys.ntu.edu.tw}}

\affiliation{${}^{1}$Tsung-Dao Lee Institute, and School of Physics and Astronomy, Shanghai Jiao Tong University, Shanghai 200240, China}
\affiliation{${}^{2}$Department of Physics, National Taiwan University, Taipei 10617, Taiwan}
\affiliation{${}^{3}$Physics Division, National Center for Theoretical Sciences, Hsinchu 30013, Taiwan}

\begin{abstract}
Among many possibilities, solar axion has been proposed to explain the electronic recoil events excess observed by Xenon1T collaboration, although it has tension with astrophysical observations. The axion couplings, to photon $g_{a\gamma}$ and to electron  $g_{ae}$ play important roles. These couplings are related to the Peccei-Quinn (PQ) charges $X_f$ for fermions. In most of the calculations, $g_{a\gamma}$ is obtained by normalizing to the ratio of electromagnetic anomaly factor $E = TrX_f Q^2_f N_c$ ($N_c$ is 3 and 1 for quarks and charged leptons respectively) and QCD anomaly factor $N = TrX_q T(q)$ ($T(q)$ is quarks'  $SU(3)_c$ index). The broken PQ symmetry generator is used in the calculation which does not seem to extract out the components of broken generator in the axion which are ``eaten'' by the $Z$ boson. However, using the physical components of axion or the ratio of anomaly factors should obtain the same results in the DFSZ for $g_{a\gamma}$. When going beyond the standard DFSZ models, such as variant DFSZ models, where more Higgs doublets and fermions have different PQ charges, one may wonder if the results are different. We show that the two methods obtain the same results as expected, but the axion couplings to quarks and leptons $g_{af}$ (here f indicates one of the fermions in the SM) are more conveniently calculated in the physical axion basis. The result depends on the values of the vacuum expectation values leading to a wider parameter space for $g_{af}$  in beyond the standard DFSZ axion. We also show explicitly how flavor conserving $g_{af}$ couplings can be maintained when there are more than one Higgs doublets couple to the up and down fermion sectors  in variant DFSZ models at tree level, and how flavor violating couplings can arise.
\end{abstract}

\maketitle

Xenon1T collaboration has observed electronic recoil events excess at energy lower than 7 kev compared with known background~\cite{xenon1t} and also compared their results with solar axion~\cite{PQ,axion,axion-dfsz,axion-ksvz}, anomalous neutrino magnetic dipole moment~\cite{neutrino}, and several models. Although the significance is only at 3.5 $\sigma$, a lot of efforts have been made to explain the excess. One should understand the background better and have more data to confirm the excess from experimental side. On the theoretical side, among many possibilities solar axion  has been proposed to explain the excess although it has tension with astrophysical observations~\cite{xenon1t, tension}. It has been shown that the inclusion of inverse Primakoff effect can significantly reduce the tension~\cite{easy-tension}. Implications of axion-like particles have also been discussed in the literature~\cite{arXiv10035, arXiv10735}. The axion couplings, to photon $g_{a\gamma}$ and to electron $g_{ae}$ play important roles.  The couplings are related to the Peccei-Quinn (PQ) charges~\cite{PQ} $X_f$ for fermions. In most of the calculations, $g_{a\gamma}$ is obtained by normalizing to the ratio of electromagnetic anomaly factor $E = TrX_f Q^2_f N_c$ ($N_c$ is 3 and 1 for quarks and charged leptons respectively) and QCD anomaly factor $N = TrX_q T(q)$ ($T(q)$ is quarks'  $SU(3)_c$ index)~\cite{kaplan,srednicki, geng}. The broken PQ symmetry generator is directly used in the calculation of the anomaly factors.

In axion models, at classical level not only the PQ global symmetry is broken by Higgs vacuum expectation values in the potential, there may be other symmetries which are also broken, such as the $SU(2)_L\times U(1)_Y$ breaking down to $U(1)_{em}$. There are at least two neutral Goldstone bosons, one corresponds to the axion and another the would-be Goldstone boson $z$ ``eaten'' by the $Z$ boson. The Goldstone boson $A$ corresponding to the broken PQ generator may not be orthogonal to $z$. The component ``eaten'' by Z boson must be extracted since that part cannot contribute to $\gamma\gamma$ due to Landau-Yang theorem~\cite{landau-yang}. The non-physical components in $A$ should be removed to obtain the physical axion $a$. This identification cannot be done before electroweak symmetry breaking because the generators of $z$ and $A$ are still symmetric and cannot be singled out. After electroweak symmetry breaking, the $z$ is ``eaten'' by Z boson to provide its longitudinal components. The physical components of $z$ are fixed. The physical axion generator must be a linear combination of the original generators for $A$ and $z$ so that the resultant one is orthogonal to $z$. One wonders if using the physical basis or the ratio of anomaly factors to calculate $g_{a\gamma}$ would obtain the same results. We take the Dine-Fischler-Srednicki-Zhitnitskii (DFSZ) models~\cite{axion-dfsz}  as example to address this question. We find that in the standard  DFSZ models, the two approaches described above give the same results. We show that when going beyond the standard DFSZ models, variant DFSZ models, even there are more Higgs bosons involved, the two approaches also give the same results, but axion couplings to quarks and charged leptons $g_{af}$ (here f indicates one of the fermions in the SM) can be very different leading to  a wider parameter space for $g_{ae}$ and in general $g_{af}$ for phenomenological studies.
We also show explicitly how flavor conserving $g_{af}$ can be maintained and flavor violating $g_{af}$ can be generated in renormalizable DFSZ and its variant models at tree level  even if there are more than one Higgs doublets couple to the up and down fermion sectors.

The standard DFSZ model has two Higgs doublets $H_1$, $H_2$ and a singlet $S$ with PQ charges $X_1=-1$, $X_2= +1$ and $X_s = -X_1 + X_2$.
\begin{eqnarray}
H_i = \left ( \begin{array} {c}
h^+_i\\
\\
{1\over \sqrt{2}}(v_i + h_i + i I_i)
\end{array}
\right ),\;\;S = {1\over \sqrt{2}}(v_s + h_s + i I_s).
\end{eqnarray}
The quarks and leptons have PQ charges
\begin{eqnarray}
Q_L:\; 0,\;\;U_R:\; X_u=X_1,\;\;D_R:\; X_d = -X_2,\;\;L_L:\;0,\;\;E_R:\; X_e= -X_2\;.
\end{eqnarray}
One can also assign $X_e = - X_1$ in the above. We will refer to these two choices of $X_e$ as standard DFSZ-I and DFSZ-II models. It is understood that there are three generations of quarks and leptons.

The PQ invariant Yukawa interaction for DFSZ-I is
\begin{eqnarray}
L_Y = - \bar Q_L Y_u \tilde H_1 U_R - \bar Q_L Y_d H_2D_R - \bar L_L Y_e H_2 E_R + H.C. \label{yukawa1}
\end{eqnarray}
where $\tilde H_i = i\sigma_2 H^*_i$.  $X_u + X_d=X_1 -X_2\neq 0$ is required for solving the strong CP problem.
For DFSZ-II, one changes $\bar L_L H_2 E_R$ to $\bar L_L H_1 E_R$.

After electroweak symmetry breaking in both models in the $(I_1,\; I_2,\; I_s)$ basis, the $z$ and $A$ are
\begin{eqnarray}
z: (v_1,\; v_2,\; 0),\;\; A: (X_1 v_1,\;X_2 v_2,\;X_s v_s).
\end{eqnarray}
As mentioned before that $A$ is not the physical axion field. The physical axion must be orthogonal component to $z$. The
orthogonal physical axion $a$ will be a linear combination of $z $ and $A$,
$a: \alpha z + A$. $\alpha$ is determined by $z\cdot a=0$ with
\begin{eqnarray}
\alpha = -{1\over v^2}(X_1 v^2_1 + X_2 v_2^2),\;\;v^2 = v^2_1 + v^2_2\;.
\end{eqnarray}
Therefore, $a \sim ( -(X_2-X_1) v^2_2v_1, (X_2-X_1)v^2_1 v_2, (X_2-X_1) v^2v_s)$.

Normalizing the field properly, we find the $z$, the axion a and the additional physical state pseudoscalar $p$ are composed of $I_{1, 2, s}$ as
\begin{eqnarray}
\left (\begin{array}{l}
z\\
\\
a\\
\\
p
\end{array}
\right ) =
\left (\begin{array}{ccc}
{v_1\over v}&\;\;{v_2\over v}&\;\;0\\
\\
{v^2_2 v_1\over v\sqrt{v_1^2v_2^2 + v^2v^2_s}}&\;\;-{v^2_1 v_2\over v\sqrt{v_1^2v_2^2 + v^2v^2_s}}&\;\;-{v^2 v_s\over v\sqrt{v_1^2v_2^2 + v^2v^2_s}}\\
\\
{v_2 v_s\over \sqrt{v^2v_s^2 + v^2_1v^2_2}}&\;\;-{v_1 v_s\over \sqrt{v^2v_s^2 + v^2_1v^2_2}}&\;\;{v_1 v_2\over \sqrt{v^2v_s^2 + v^2_1v^2_2}}
\end{array}
\right )\left (\begin{array}{l}
I_1\\
\\
I_2\\
\\
I_s
\end{array}
\right ).
\end{eqnarray}

The tree level Yukawa coupling of axion for DFSZ-I is given by
\begin{eqnarray}
L_{Y-a} = i{a\over v\sqrt{v^2_1v^2_2 + v^2 v^2_s}}( v^2_1 \bar U M_u \gamma_5 U + v^2_2\bar D M_d \gamma_5 D + v^2_2 \bar E M_e \gamma_5 E)\;.
\end{eqnarray}
Replacing $v_2^2$ by $- v_1^2$ in front of  $\bar E M_e \gamma_5 E$ for the electron coupling, one can obtain the axion coupling to fermions in DFSZ-II.

We now use physical axion $a$ to calculate the well known triangle diagrams for
the axion-gluon and axion-photon interactions. We obtain
\begin{eqnarray}
&&L_{agg} = N {g^2_3\over 16 \pi^2} {1\over v\sqrt{v^2_1v^2_2 + v^2v_s^2}}(v^2_1+v^2_2) a T(q)G^a_{\mu\nu} \tilde G^{\mu\nu}_a =
{\alpha_s \over 8 \pi }{a\over f_a}G^a_{\mu\nu} \tilde G^{\mu\nu}_a \;,\nonumber\\
&&L_{a\gamma\gamma} = N {e^2\over 16 \pi^2} {1\over v\sqrt{v^2_1v^2_2 + v^2v_s^2}}((v^2_1Q_u^2+v^2_2Q^2_d)N_c + v^2_2 Q_e^2) a F_{\mu\nu} \tilde F^{\mu\nu} = {1\over 4} a g^0_{a\gamma} F_{\mu\nu} \tilde F^{\mu\nu}\;,
\end{eqnarray}
where $N=3$ is the generation number, and $N_c = 3$ is the number of color. $T(q)$ is the quark $SU(3)_C$ index defined by $Tr(T^aT^b)=T(q)\delta^{ab}= (1/2)\delta^{ab}$ since quark is a fundamental representation of $SU(3)_C$. One can read off the axion decay constant $f_a$ and the axion-photon coupling $g_{a\gamma}^0$ as
\begin{eqnarray}
{1\over f_a } = {2 N v  \over \sqrt{v^2_1v^2_2 + v^2 v_s^2}}  T(q), \;\;g^0_{a\gamma} = {\alpha_{em}\over 2 \pi f_a} {E(\tilde X)\over N(\tilde X)}\;.
\end{eqnarray}
Here $N(\tilde X) = \sum_{i = u, d} \tilde X_i  T(q)$ and $E(\tilde X) = \sum_{i=u,d,e} \tilde X_i Q^2_i N_c^i$. Here $N_c^i$ is 3 for quarks and 1 for leptons, and
\begin{eqnarray}
\tilde X_u = {v_1^2\over v^2_1+v_2^2},\;\;\tilde X_d = {v_2^2\over v^2_1+v_2^2},\;\;\tilde X_e = {v_2^2\over v^2_1+v_2^2},
\end{eqnarray}
for DFSZ-I. For DFSZ-II couplings, one just replaces $\tilde X_e = v_2^2/ (v^2_1+v_2^2)$ by $\tilde X_e = - v_1^2/ (v^2_1+v_2^2)$.
If $v_s >> v$, $f_a \approx v_s/(2NT(q))$.

The axion couplings to light fermions $u,\;d,\;s$ and $e$ can be straightforwardly written into derivative form
\begin{eqnarray}
L_{Y-a} = -{\partial_\mu a \over 2 f_a} {1\over N} \sum_{q=u,d,e} \bar q \tilde X_q \gamma^\mu \gamma_5 q = {\partial_\mu a \over 2 f_a} j^\mu_{a,0}\;. \label{yukawa}
\end{eqnarray}
For DFSZ-I, it is equal to
\begin{eqnarray}
L_{Y-a} =- {1\over 2} {\partial_\mu a\over v\sqrt{v^2_1v^2_2 + v^2 v^2_s}}(  v^2_1 \bar u \gamma^\mu \gamma_5 u + v^2_2\bar d \gamma^\mu \gamma_5 d + v^2_2 \bar e \gamma^\mu \gamma_5 e).
\end{eqnarray}
Again for DFSZ-II couplings, one just replaces $v^2_2$ to $-v^2_1$ in front of the term involving electron e.

We have obtained the axion couplings in the physical axion basis. Concerning the axion-photon coupling, if one calculates the axion-photon coupling $g^0_{a\gamma}$ using $A$, as did by many previously, one would obtain a similar expression by replacing $X_i$ by $\tilde X_i$  respectively to obtain $N( X)$ and $E(X)$~\cite{kaplan, srednicki,geng}.
We now check if the ratio of $E$ and $N$ is equal in both basis. We have
\begin{eqnarray}
\mbox{For DFSZ-I}:\;\;\;\;&&{E(X)\over N(X) }=
{ (X_u Q^2_u + X_d Q^2_d)N^q_c + X_e Q^2_eN_c^e\over  (X_u+X_d)T(q)} = {4/3+1/3 + 1 \over (1+1)(1/2)} = {8\over 3} \;,\nonumber\\
&&{E(\tilde X)\over N(\tilde X) }=
{ (\tilde X_u Q^2_u + \tilde X_d Q^2_d)N^q_c + \tilde X_e Q^2_eN_c^e\over  (\tilde X_u+\tilde X_d)T(q)} = {4/3\times v^2_1+1/3\times v^2_2 + 1\times v^2_2 \over (1\times v^2_1+1\times v^2_2)(1/2)} = {8\over 3}\;.\nonumber\\
\mbox{For DFSZ-II}:\;\;&&{E(X)\over N(X) }=
{ (X_u Q^2_u + X_d Q^2_d)N^q_c + X_e Q^2_eN_c^e\over  (X_u+X_d)T(q)} = {4/3+1/3 - 1 \over (1+1)(1/2)} = {2\over 3} \;,\\
&&{E(\tilde X)\over N(\tilde X) }=
{ (\tilde X_u Q^2_u + \tilde X_d Q^2_d)N^q_c + \tilde X_e Q^2_eN_c^e\over  (\tilde X_u+\tilde X_d)T(q)} = {4/3\times v^2_1+1/3\times v^2_2 - 1\times v^2_1 \over (1\times v^2_1+1\times v^2_2)(1/2)} = {2\over 3}\;.\nonumber
\end{eqnarray}
The above two models obtain the same ratio $E/N$ for the two methods.

The advantages of using physical axion for calculations are that one can obtain directly the axion-fermion couplings without the need of further manipulations to extract the components which are ``eaten'' by the Z boson as can be seen in the above calculations.

Can one draw a conclusion that the two methods described above always give the same results? To answer this question, let us consider a variant DFSZ model in which the PQ charge of charged leptons $E_R$ is set to be 0 so that it couples to a different Higgs doublet $H_3$ with a PQ charge to be $0$\cite{geng}. We refer this model as vDFSZ-I model. In this case the term $\bar L_L Y_e H_2 E_R$ in eq(\ref{yukawa1}) is replaced by $\bar L_L Y_e H_3 E_R$. The PQ charges for other fields do not change as those in the standard DFSZ model.

Because there are more Higgs doublets with different PQ charges in vDFSZ-I model, there may be some additional global symmetries in the Higgs potential to have additional Goldstone boson after electroweak symmetry breaking which complicates the analysis. To avoid this to happen, we assign the singlet  $S$ to have a PQ charge $X_s = -1$. The renormalizable Higgs potential admits terms, like, $H^\dagger_3 SH_2$, $H^\dagger_1 S H_3$ and $H^\dagger_1 S^2 H_2$. Therefore the potential does not have additional global symmetry except PQ  and  the $SU(2)_L\times U(1)_Y$ gauge symmetries in the standard DFSZ models. After electroweak symmetry breaking, the two broken generators  corresponding to $z$ and $A$ are $z: \;(v_1,\;v_2,\;v_3,0)$ and $A:\;(X_1 v_1,\;X_2v_2,\;0, X_s v_s)$. Following the same procedure before, we obtain the physical axion field to be  given by
\begin{eqnarray}
a = {1\over N_a} \left ( (2v^2_2+v^2_3)v_1I_1 - (2v^2_1 + v_3^2) v_2 I_2 - (v_1^2-v_2^2) v_3 I_3 + v^2v_s I_s \right )\;.
\end{eqnarray}
where $N_a^2 = ((2v^2_2 + v_3^2) v_1)^2 + ((2v^2_1 + v_3^2) v_2)^2 + ((v_1^2-v_2^2)v_3)^2 + (v^2v_s)^2$  is a normalization constant with $v^2 = v^2_1 +v^2_2+v^2_3$.
One obtains
\begin{eqnarray}
L_{Y-a} = i{a\over N_a}( (2v^2_2 + v_3^2) \bar U M_u \gamma_5 U + (2v^2_1 + v_3^2) \bar D M_d \gamma_5 D + (v_1^2-v_2^2) \bar E M_e \gamma_5 E) = - {\partial_\mu a \over 2 f_a} {1\over N} \sum_{q=u,d,e} \bar q \tilde X_q \gamma^\mu \gamma_5 q\;,
\end{eqnarray}
where the axion decay constant is now given by $f_a^{-1} = 4N v^2T(q)/N_a$ and
\begin{eqnarray}
\tilde X_u = {2 v^2_2 + v^2_3\over v^2},\;\;\;\;\tilde X_d = {2 v^2_1 + v^2_3\over v^2},\;\;\;\;\tilde X_e = {v^2_1 - v^2_2\over v^2}.
\end{eqnarray}

Carrying out the one loop triangle diagram calculations, we would obtain
\begin{eqnarray}
{E(\tilde X)\over N(\tilde X)} =  { ((2v_2^2+ v_3^2)Q_u^2 + (2v_1^2+v_3^2)Q_d^2)N^q_c + (v^2_1 - v_2^2)Q_e^2
\over ((2 v_2^2+v_3^2) + (2v_1^2+v_3^2))T(q)} = {4(2v^2_2 + v^2_3) +(2v^2_1+v^2_3) + 3(v^2_1-v^2_2)\over 3 (v^2_1+v^2_2+v^2_3)} = {5\over 3}\;.
\end{eqnarray}
while $E(X)/N(X)$ would give
\begin{eqnarray}
{E(X)\over N(X)} = {(X_uQ^2_u + X_dQ_d^2)N^q_c \over (X_u + X_d)T(q)} = {5\over 3}\;.
\end{eqnarray}
The ratios $E(\tilde X)/N(\tilde X)$ and $E(X)/N(X)$ are obviously equal to each other for the two ways of calculating the $g_{a\gamma}$.

One can even make the three Higgs doublets to have different PQ charges, $H_1: X_1 = -X_u$, $H_2: X_2 = X_d$ and $H_3: X_3 = X_e$. This is the vDFSZ-II model to be discussed. The three terms can exist $H^\dagger_1 S H_2$, $H^\dagger_1  S^2 H_3$ and $H^\dagger_3 S^\dagger H_2$  for any $X_e$ which determine $X_s = -(X_u + X_d)$. We would have
\begin{eqnarray}
a = {1\over N_a} (&-&[(X_u+X_d)v^2_2 +(X_u + X_e)v^2_3]v_1 I_1 + [(X_u+X_d)v^2_1
+(X_d - X_e)v^2_3]v_2 I_2\nonumber\\
&+&[(X_e + X_u) v^2_1 + (X_e - X_d)v^2_2]v_3 I_3 + X_s v^2 v_sI_s)\;.
\end{eqnarray}
Denoting
\begin{eqnarray}
\tilde X_u = {(X_u+X_d)v^2_2 +(X_u + X_e)v^2_3\over v^2},\;
\tilde X_d = {(X_u+X_d)v^2_1 +(X_d- X_e)v^2_3\over v^2},\;
\tilde X_e = {(X_e+X_u)v^2_1 +(X_e-  X_d)v^2_2\over v^2},
\end{eqnarray}
we have
\begin{eqnarray}
{E(\tilde X)\over N(\tilde X) } = {(X_u Q^2_u + X_d Q^2_d)N^q_c
+ X_e Q^2_eN^e_c \over (X_u + X_d)T(q)} = {E(X)\over N(X)}\;.
\end{eqnarray}
This is a more general proof that the two approaches give the same result.
The axion-fermion interaction has the same form as that of eq. (\ref{yukawa}) with appropriate use of $\tilde X_i$ for this model.

We now discuss the possibility that whether flavor changing interaction of an axion exists in a renormalizable variant DFSZ models. For multi Higgs doublet models~\cite{glashow-weinberg}, it is possible to have flavor changing interactions of a neutral scalar/pseudoscalar with fermions. In fact this happens for two Higgs doublet models in general because both Higgs doublets can have Yukawa couplings to the up or down sectors. In the standard DFSZ models, there are also two Higgs doublets. However, because they have different PQ charges, only one of them can couple to the up or down sector separately. There is no flavor changing interactions for neutral scalar/pseudoscalar in the models at tree level.  We therefore need to introduce more Higgs doublets so that at least two Higgs doublets can couple to the up or down sector. To this end we take an axion model, vDFSZ-III model, discussed in ref. \cite{phase} where there are three Higgs doublets $H_{1, 2}: (1,2, -1/2) (+1)$, $H_3: (1,2,-1/2) (-1)$ and a singlet $S: (1,1,0) (2)$. Here the quantum numbers in the first bracket are for $SU(3)_c\times SU(2)_L\times U(1)_Y$ ones and the numbers in the second bracket are the PQ charges.

The Yukawa couplings in this model are given by
\begin{eqnarray}
L_Y = -\bar Q_L Y_u H_3 U_R - \bar Q_L (Y_{d1} \tilde H_1 + Y_{d_2} \tilde H_2) D_R -
\bar L_L (Y_{e1} \tilde H_1 + Y_{e_2} \tilde H_2) E_R + H.C..
\end{eqnarray}
Since there are two Higgs doublets which couple to the down sectors for quarks and leptons, there should be in general flavor changing interactions of neutral scalar/pseudoscalar bosons to down quarks and also charged leptons.

We now show that the axion will remain to have flavor conserving interactions with quarks and charged leptons at the tree level. The axion field in this model is given by
\begin{eqnarray}
a &=& {1\over v \tilde N_a} (v_1 v^2_3 I_1 + v_2 v^2_3 I_2 - v_3 (v^2_1+v_2^2) I_3 - v_sv^2 I_s)\nonumber\\
&=&{v\over \tilde N_a}(\tilde X_1 v_1 I_1 + \tilde X_2 v_2 I_2 + \tilde X_3 v_3 I_3 + \tilde X_s v_s I_s)\;.
\end{eqnarray}
where $v^2= v^2_1 + v^2_2 + v^2_3$ and $\tilde N_a^2 = (v^2_1+v^2_2)v^2_3 + v^2_s v^2$,
and
\begin{eqnarray}
\tilde X_1 = {v^2_3\over v^2}=\tilde X_2 = {v^2_3\over v^2}= \tilde X_d =\tilde X_e\;,\;\;\;\;\tilde X_3 = {v^2_1+v^2_2\over v^2}=\tilde X_u\;, \;\;\;\;
\tilde X_s = {v^2\over v^2}=1\;.
\end{eqnarray}
This gives the axion couplings to fermions as the following
\begin{eqnarray}
L_{a-Y} = -i{a\over v \tilde N_a} \left ((v^2_1+v^2_2)  \bar U_L {Y_uv_3\over \sqrt{2}} U_R +v^2_3 \bar D_L ({Y_{d1} v_1\over \sqrt{2}} + {Y_{d_2} v_2\over \sqrt{2}}) D_R +
v^2_3\bar E_L ({Y_{e1} v_1\over \sqrt{2}} + {Y_{e_2}v_2\over \sqrt{2}} ) E_R \right )+ H.C..
\end{eqnarray}

Since the Yukawa couplings $Y_i$ are in general not diagonal, naively, one might expect that axion $a$ has flavor changing interactions with quarks and leptons. However, note that in this model the mass matrices for up and down quarks and charged leptons are
\begin{eqnarray}
M_{u} = {Y_u v_3\over \sqrt{2}},\;\;M_{d} = {Y_{d1} v_1\over \sqrt{2}} + {Y_{d2} v_2\over \sqrt{2}} , \;\;M_{e} = {Y_{e1} v_1\over \sqrt{2}} + {Y_{e2} v_2\over \sqrt{2}} .
\end{eqnarray}
The above axion-Yukawa couplings are all proportional to the mass matrices. In the basis where the mass matrices $M_i$ are diagonalized to $\hat M_i$, the axion Yukawa interactions are also diagonal with
\begin{eqnarray}
L_{a-Y} = -ia {v\over \tilde N_a} \left (\tilde X_u  \bar U \hat M_u\gamma_5 U +\tilde X_d \bar D \hat M_d \gamma_5 D +
\tilde X_e \bar E\hat M_e \gamma_5 E \right )\;.
\end{eqnarray}
Therefore we see that there is no flavor changing interaction between axion and quarks and leptons  in this model. Using the above axion-fermion interaction, one can obtain in a similar way the axion-photon coupling. Note that for each up and down sectors, $\tilde X_i$ are the same and therefore can be factored out in front of the mass matrices parts. One also obtains $E(\tilde X)/N(\tilde X) = E(X)/N(X)$.

The axion-Yukawa coupling matrices and the mass matrices can be simultaneously diagonalized in axion models is the key to eliminate flavor changing interactions of the axion to fermions. It happens in axion models is due to the fact that the physical axion field always comes with the combination$\tilde X_f Y_i v_i I_i $ which leads to axion couplings to fermions to be proportional to  $\tilde X_f \bar f Y_i v_i \gamma_5 f a$ for the i-th Higgs doublet for a particular up or down sector of axion-fermion interaction $\sim \sum Y_i v_i/\sqrt{2}$, which in turn contributes to the $Y_i v_i$ component of the total mass matrix $M = \sum Y_i v_i/\sqrt{2}$.  One of the crucial condition here is that the $\tilde X_i$ for each up and down sectors for quarks and charged leptons have the same $\tilde X_i$. The neutral scalar bosons, and the orthogonal component physical pseudoscalar bosons to axion can have flavor changing interactions, but axion couplings to fermion are flavor conserving at the tree level in the vDFSZ-III model. At loop levels, flavor changing interactions of axion with fermions can be generated, just like in the SM, flavor changing interaction of Higgs is generated at loop levels.

So far we have been concentrating on possible renormalizable sources of flavor changing axion-fermion interactions. However, if there are non-renormalizable axion-fermion interactions, for example, $\bar Q_L (\tilde Y_u/\Lambda^2) H_3U_R  (H_3^\dagger H_3)$, when the renormalizable contributions to the  mass matrices are diagonalized as before, $\tilde Y_u$ may not be simultaneously diagonalized. This non-renormalizable terms can generate flavor changing axion-fermion interactions. In a renormalizable axion models, axion does not generate tree level flavor changing interactions in the above type of models where the resulting $\tilde X_i$ for the up and down sectors can be factored out.

One way to generate flavor violating axion-fermion couplings is to let the $\tilde X_i$ not to be factored out for each up or down sectors. It has been shown this is possible with three Higgs doublets variant DFSZ models~\cite{arXiv6217, arXiv6218}. Here, we demonstrate this by a specific model with different PQ charges for different generations in the down quarks sector, vDFSZ-IV model,  following the same procedure to identify axion field and then obtain the axion-fermion couplings.  The model has 3 Higgs doublets with PQ charges: $H_1:(X_1=-1)$, $H_2: (X_2= - 2)$, $H_3: (X_3=-3)$, and a singlet with PQ charge $S:(X_s=1)$. The fermions have PQ charges, $L^i_{L }: (X_L = 0)$, $E^i_{ R }: (X_R = 1)$, $Q^i_{L }: (X_Q=0)$, $U^i_{R }: (X_u = -1)$, $D_R^1: (X_d^1=1)$, $D_R^2: (X_d^2=2)$ and $D_R^3: (X_d^3=3)$. The relevant Yukawa coupling interaction is given by
\begin{eqnarray}
L_Y = - \bar L_L Y_l H_1 E_R - \bar Q_{L} Y_u \tilde H_1 U_R - \bar Q_{L i} Y^{i1}_d H_1D_R^1 - \bar Q_{L i} Y^{i2}_d H_2D_R^2- \bar Q_{L i} Y^{i3}_d H_3D_R^3 + H.C.. \label{model2}
\end{eqnarray}

The physical axion field is given by
\begin{eqnarray}
a = {1\over N'_a} ( \sum_i (X_i - X_l) v^2_i v_l I_l - v^2 X_s v_s I_s)\;,
\end{eqnarray}
where $v^2 = v_1^2+v_2^2+v^2_3$ and  {$N'_a$ represents a new normalization constant.

In the Yukawa interaction of Eq.(\ref{model2}), the down quarks interact with more than one Higgs doublets and also different generations have different PQ charges, they have the potential to generate flavor changing axion coupling for down type quarks. We will concentrate on this interaction. We obtain the down quark mass matrix $\bar D_L M_d D_R$ and axion coupling matrix $ i \bar D_L C_d D_R$ as the following
\begin{eqnarray}
M_d = {1\over \sqrt{2}} \left ( \begin{array}{ccc}
Y^{11}_d v_1\;\;\; & Y^{12}_d v_2 \;\;\;&Y^{13}_d v_3\\
\\
Y^{21}_d v_1\;\;\; & Y^{22}_d v_2 \;\;\;&Y^{23}_d v_3\\
\\
Y^{31}_d v_1\;\;\; & Y^{32}_d v_2 \;\;\;&Y^{33}_d v_3
\end{array}
\right )\;,\;\;\;\;
C_d = {v^2 \over N'_a} M_d \tilde X\;,
\end{eqnarray}
where $\tilde X = diag(\tilde X_1,\;\tilde X_2,\;\tilde X_3)$ is a diagonal matrix with
\begin{eqnarray}
\tilde X_1 =  -{ v_2^2 + 2 v^2_3 \over v^2}\;,\;\;\;\;\tilde X_2 = {v^2_1 - v^2_3 \over v^2}\;,\;\;\;\; \tilde X_3 = {2 v^2_1 + v^2_2\over v^2}\;.
\end{eqnarray}

$M_d$ can be diagonalized to obtain the diagonal mass matrix $\hat M_d$ by bi-unitary transformation $\hat M_d = V^\dagger_L M_d V_R$. In the mass eigen-basis,
the coupling matrix becomes $V^\dagger_L C_d V_R = (v^2/N'_a)\hat M_d V^\dagger_R \tilde X V_R$. $M_d$ and $C_d$ cannot be diagonalized simultaneously if $\tilde X$ is not proportional to a unity matrix as it is in our model. This leads to flavor violating axion-down quark couplings.
We therefore  have shown that in vDFSZ-IV model axion-fermion couplings have flavor violating terms. For phenomenological studies of flavor changing interactions in such models, one
should take into possible contributions from axion interactions along with other possible scalar/pseudoscalar particles to have a complete pictures.
The detailed axion-fermion couplings in the model above are different from those in models discussed in ref.\cite{arXiv6217, arXiv6218} and can in principle be tested once axion is observed. As far as generating flavor changing axion-fermion interactions, they all serve as concrete examples.

As a check we also calculate the ratio $E(X)/N(X)$ and the $E(\tilde X)/N(\tilde X)$.   In doing the calculation, one needs to examine each individual generation of fermions as the down quarks do not have a universal PQ charge $X_d^i$ for the three generations and also the corresponding $\tilde X_d^i$. We obtain the same result
with $E(X)/N(X)=E(\tilde X)/N(\tilde X)= 2/3$.

In the above we have discussed several different types of DFSZ and its variant axion models, the standard DFSZ-(I, II) and vDFSZ-(I, II, III, IV). We have shown that the axion couplings to quarks, leptons, gluons and photons can be easily calculated in the physical axion basis with would-be Goldstone boson `eaten' by Z removed. Different assignments of PQ charges have different consequences for axion couplings to other particles.
The different PQ charges for various models also lead to different cosmological implications. A closely related one is the potential domain wall problem with axion models where different PQ charge assignments determine the domain wall number.  Here we briefly comment on the domain wall numbers in each model discussed.

DFSZ axion model and its variants have spontaneously broken discrete symmetries depending on the PQ charge of the singlet scalar which sets the PQ symmetry breaking scale and also the PQ charges of quarks~\cite{arXiv6217, arXiv6218, geng11}.  The instanton effects of QCD will break chiral $U(1)_A$ down to discrete $Z_{2N_f}$ symmetry due to gluon anomaly $SU(3)_L^2\times U(1)_{PQ}$ and be reduced further by the singlet vaccum property related to its PQ charge. The models finally arrive at $N_{DW}$ number of degenerate disconnected vacua~\cite{sikivie}, which leads to some unwanted domain wall structure in the early universe~\cite{dw-problem} if the domain wall number $N_{DW}$ is not equal to 1. The number $N_{DW}$ is determined by~\cite{geng11} $\sum_i (2 X^i_{Q} + X^i_{u}+ X^i_{d})/X_s$. We obtain the following domain wall numbers for the models discussed earlier,
\begin{eqnarray}
&Model&:\;\;       \mbox{DFSZ-I} \;\;\;\;   \mbox{DFSZ-II}\;\;\;\; \mbox{vDFSZ-I}\;\;\;\; \mbox{vDFSZ-II}\;\;\;\; \mbox{vDFSZ-III}\;\;\;\;\mbox{vDFSZ-IV}\nonumber\\
&N_{DW}&:\;\;\;\;\;\;\;\;        3 \hspace{1.4cm}   3 \hspace{1.5cm}    6\hspace{1.7cm}    3\hspace{1.7cm}   3\hspace{1.7cm}    3
\end{eqnarray}

All of the above models have $N_{DW}$ larger than $1$ and have potential domain wall problem. This problem can be remedied by assuming the PQ symmetry breaking happened before inflation so that harmful effects of domain wall were washed out. One may also try to modify the PQ charge assignments to have $N_{DW}=1$. This can be achieved by assigning
the three up right-handed quarks to have $X^1_u = -1$, $X^2_u = -1$, $X^3_u = -3$ instead all have $X_u = -1$ as in vDFSZ-IV model. In this case, the model is free from domain wall problem and at the same time has flavor changing axion-fermion interactions.

For completeness,  before conclusion, let us briefly outline how to match below electroweak scale the quark level  axion couplings to hadrons using effective interactions below electroweak scale obtained. We summarize them in the following for the interactions of axion with light fermions $q = (u,\;d,\;s)$ and $e$, gluon and photon~\cite{georgi,new-axion},
\begin{eqnarray}
L_a = {1\over 2} \partial^\mu a \;\partial_\mu a + {a\over f_a} {\alpha_s\over 8 \pi} G^a_{\mu\nu}\tilde G^{\mu\nu}_a+ {1\over 4} a\; g^0_{a\gamma} F_{\mu\nu}\tilde F^{\mu\nu} + {\partial_\mu a \over 2 f_a} j^\mu_{a,0} - (\bar q_L  M_q q_R + H.C.)
\end{eqnarray}
where $M_q$ is a diagonal mass matrix with diagonal entries: $(m_u,\;m_d,\;m_s)$.

Our discussions follow closely ref.\cite{georgi}. The axion-gluon coupling can be explicitly removed from the effective Lagrangian by performing a chiral rotation on the quarks as the following coupling form
\begin{eqnarray}
q \to e^{-i\gamma_5 {a\over 2 f_a}Q_a} q,
\end{eqnarray}
one obtains
\begin{eqnarray}
L_a = {1\over 2} \partial^\mu a \;\partial_\mu a + {1\over 4} a\; g_{a\gamma} F_{\mu\nu}\tilde F^{\mu\nu} + {\partial_\mu a \over 2 f_a} j^\mu_{a} - (\bar q_L  M_a q_R + H.C.)
\end{eqnarray}
with
\begin{eqnarray}
g_{a\gamma} = {\alpha_{em}\over 2 \pi f_a} \left ({E(\tilde X)\over N(\tilde X)} - 6 Tr(Q_a Q^2)\right )\;,\;\;\;\;
j^\mu_a = j^\mu_{a,0} - \bar q \gamma^\mu \gamma_5 Q_a q\;,\;\;\;\; M_a = e^{i{a\over 2 f_a} Q_a} M_q e^{i{a\over 2 f_a} Q_a}\;.
\end{eqnarray}
where $Q_a$ is chosen to be a diagonal one with the entries to be $(Q^{11}_a,\;Q^{22}_a,\;Q^{33}_a)=(1/m_u,\; 1/m_d,\;1/m_s)/(1/m_u + 1/m_d + 1/m_s)$ so that there will be no axion-$\pi^0$ mixing. Other choice is also allowed, but there will be axion-$\pi^0$ mixing which requires further diagonalization, but at the end one will obtain the same results.

The chiral realization to obtain the axion mass is closely related to how pions obtain their masses. The leading order meson masses are given by
\begin{eqnarray}
&&L_{a-mass}  =  2 B_0 {f^2_\pi\over 4} Tr(U M^\dagger_a + M_a U^\dagger)\;,\;\;U = e^{i\sqrt{2}\Pi/f_\pi},\nonumber\\
&&\nonumber\\
&&\Pi = \left (\begin{array}{ccc}
\pi^0/\sqrt{2}+ \eta/\sqrt{6}& \pi^+&K^+\\
\pi^-&-\pi^0\sqrt{2} +\eta/\sqrt{6}& K^0\\
K^-&\bar K^0&-2\eta/\sqrt{6}
\end{array}
\right )\;.
\end{eqnarray}
Specializing to $a$ and $\pi^0$, we obtain
\begin{eqnarray}
L_{a-mass} = B_0f^2_\pi \left (m_u \cos(\pi^0/f_\pi -Q^{11}_a\; a/f_a ) + m_d\cos(\pi^0/f_\pi + Q^{22}_a\; a /f_a) \right )\;.
\end{eqnarray}
We obtain the axion mass $m_a$ to be
\begin{eqnarray}
m_a^2 = {f^2_\pi\over f_a^2} m^2_{\pi^0} {m_u m_d m_s \over (m_u+m_d )(m_um_d + m_u m_s + m_d m_s)} \approx {f^2_{\pi}\over f^2_a} m^2_{\pi^0}{m_um_d\over (m_u+m_d)^2}\;.
\end{eqnarray}

To obtain the axion-baryons couplings, one matches the current term $j^\mu_a$ using chiral Lagrangian realization following ref.\cite{georgi}. One obtains~\cite{kaplan, georgi}
\begin{eqnarray}
L_{a-N}={\partial_\mu a\over 2 f_a} \left (2 Tr(({\tilde X - Q_a}) T^a)
\left (F Tr(\bar B \gamma^\mu \gamma_5[T^a, B]) + D Tr(\bar B \gamma^\mu \gamma_5 \{T^a,B\})\right )
+ {1\over 3} Tr({\tilde X - Q_a}) S Tr(\bar B \gamma^\mu \gamma_5 B)\right )\;, \nonumber\\ \label{a-N}
\end{eqnarray}
where $T^a=\lambda^a/2$ with $\lambda^a$ being the Gell-Mann matrices. $\tilde X = diag(\tilde X_u,\; \tilde X_d,\;\tilde X_s)$ are model dependent as discussed in previous sections. $D = 0.81$, $F=0.44$~\cite{df} and $S$ is between 0.0 to 2.2 ~\cite{georgi}.

Using the above one can obtain the isoscalar $g_0$ and iso-vector $g_3$ axion-nucleon couplings.
One can also use Goldberger-Treinan relation to obtain the iso-scalar $g_0$ and iso-vector $g_3$ couplings  to proton and neutron $\psi = (p, n)^T$ defined by~\cite{srednicki}
$L_{a-N} = a \bar \psi (g_0 + g_3 \tau_3)\psi$,
\begin{eqnarray}
&&g_0 = -F_{A0}\left ({1\over N}(\tilde X_u + \tilde X_d) - {1 +z\over 1+z+w}\right ) {m_N\over 2f_a}\;,\nonumber\\
&&g_3 = -F_{A3}\left ({1\over N}(\tilde X_u - \tilde X_d) - {1 - z\over 1+z+w}\right ) {m_N\over 2f_a}\;,
\end{eqnarray}
where $z=m_u/m_d$ and $w=m_u/m_s$.

Compared with the results in eq.(\ref{a-N}), we obtain
\begin{eqnarray}
F_{A0} = {1\over 3} (D-3F-2S)\;,\;\;\;\;F_{A3} = -(D+F) \;.
\end{eqnarray}

While the Xenon1T collaboration reporting their data, they also performed a solar axion fit to the data to explain the recoil electron events excess and pointed out their solar axion fit may have tension with astrophysical observations. It has also been shown that the inclusion of inverse Primakoff effect can reduce significantly the tension~\cite{easy-tension}. Ignoring the astrophysical constraints, we find that both DFSZ-I and DFSZ-II  model can explain the Xenon1T data consistently as claimed in ref.\cite{xenon1t}. For example, for the bench mark set of parameters after taking into account the inverse Primakoff effect, $g_{ae} \sim 1.5\times 10^{-13}$ and $g_{a\gamma} \sim 2\times 10^{-10}$ GeV$^{-1}$, the standard DFSZ-I model would obtain $f_a \sim 4\times 10^6$ GeV, $m_a \sim 1.3$ eV, $v_2/v \sim 0.062$ and also predict
$g_{an} \sim 6 \times 10^{-8}$. For DFSZ-II model, one would obtain $f_a \sim 7 \times 10^6$ GeV, $m_a \sim 0.79$ eV, $v_2/v \sim 0.08$ and also predict
$g_{an} \sim 3 \times 10^{-8}$.

To conclude, the axion couplings, $g_{a\gamma}$ and $g_{ae}$ to photon and electron can play important roles in explaining the Xenon1T data.
We have reexamined theoretical calculations for axion couplings to fermions and photon. The axion couplings are related to the PQ charges $X_f$ for fermions and Higgs bosons. We have confirmed the results that the axion-photon coupling $g_{a\gamma}$ obtained by normalizing to the ratio of electromagnetic anomaly factor $E$ and QCD anomaly factor $N$  and explicit calculation using physical axion obtain the same results. For calculating axion couplings to fermions, It is more convenient to use a basis where axion is already identified as the physical one. The coupling $g_{ae}$ can have very different values in different models.
We also show explicitly how flavor conserving can be maintained when there are more than one Higgs doublets coupling to the up and down fermion sectors  in variant DFSZ models at tree level, and how flavor violating axion couplings to quarks and charged leptons can arise.

XG thanks for Prof. Chao-Qiang Geng and Prof. Cheng-Wei Chiang for useful discussions.  XGH was supported in part by the MOST (Grant No. MOST 106-2112-M-002-003-MY3 ). JS was supported in part by Key Laboratory for Particle Physics, Astrophysics and Cosmology, Ministry of Education, and Shanghai Key Laboratory for Particle Physics and Cosmology (Grant No. 15DZ2272100).


\begin{thebibliography}{99}

\bibitem{xenon1t} E.~Aprile \textit{et al.} [XENON],
[arXiv:2006.09721 [hep-ex]].

\bibitem{PQ} R. D. Peccei and H. R. Quinn, CP conservation in the presence of instantons, Phys. Rev. Lett. 38 (1977) 1440. [328(1977)];
R. D. Peccei and H. R. Quinn, Constraints Imposed by CP Conservation in the Presence of Instantons, Phys. Rev. D 16 (1977) 1791.

\bibitem{axion} S. Weinberg, Phys. Rev. Lett. 40, 223 (1978); F. Wilczek, Phys. Rev. Lett. 40, 279 (1978).

\bibitem{axion-dfsz} M. Dine, W. Fischler, and M. Srednicki, Physics Letters B104, 199 (1981); A. Zhitnitskii, Sov. J. Nucl. Phys. 31, 260 (1980).

\bibitem{axion-ksvz} J. E. Kim, Phys. Rev. Lett. 43, 103 (1979); M. Shifman, A. Vainshtein, and V. Zakharov, Nuclear
Physics B 166, 493 (1980).

\bibitem{neutrino} N. F. Bell, V. Cirigliano, M. J. Ramsey-Musolf, P. Vogel, and M. B. Wise, Physical Review Letters 95, 14 (2005); N. F. Bell, M. Gorchtein, M. J. Ramsey-Musolf, P. Vogel, and P. Wang, Physics Letters, Section B: Nu- clear, Elementary Particle and High-Energy Physics 642, 377383 (2006).

\bibitem{tension} L.~Di Luzio, M.~Fedele, M.~Giannotti, F.~Mescia and E.~Nardi,
[arXiv:2006.12487 [hep-ph]].

\bibitem{easy-tension} C.~Gao, J.~Liu, L.~T.~Wang, X.~P.~Wang, W.~Xue and Y.~M.~Zhong,
[arXiv:2006.14598 [hep-ph]];
J.~B.~Dent, B.~Dutta, J.~L.~Newstead and A.~Thompson,
[arXiv:2006.15118 [hep-ph]].

\bibitem{arXiv10035}
F.~Takahashi, M.~Yamada and W.~Yin,
[arXiv:2006.10035 [hep-ph]].

\bibitem{arXiv10735}
K.~Kannike, M.~Raidal, H.~Veermäe, A.~Strumia and D.~Teresi,
[arXiv:2006.10735 [hep-ph]].



\bibitem{kaplan} D. B. Kaplan, Nuclear Physics B 260, 215226 (1985).

\bibitem{srednicki} M. Srednicki, Nuclear Physics B 260, 689700 (1985).

\bibitem{geng} S.~Cheng, C.~Geng and W.~Ni,
Phys. Rev. D \textbf{52}, 3132-3135 (1995)
doi:10.1103/PhysRevD.52.3132
[arXiv:hep-ph/9506295 [hep-ph]].

\bibitem{landau-yang}  Lev Davidovich Landau
Dokl. Akad. Nauk SSSR. 60: 207?209 (1948).
Chen Ning Yang,
Physical Review 77, 242?245 (1950).

\bibitem{glashow-weinberg} S.~L.~Glashow and S.~Weinberg,
Phys. Rev. D \textbf{15}, 1958 (1977)
doi:10.1103/PhysRevD.15.1958.

\bibitem{phase} J.~Pan, J.~Sun, X.~D.~Ma and X.~G.~He,
Phys. Lett. B \textbf{807}, 135573 (2020)
doi:10.1016/j.physletb.2020.135573
[arXiv:2003.09921 [hep-ph]].

\bibitem{arXiv6217}
A.~Celis, J.~Fuentes-Martin and H.~Serodio,
Phys. Lett. B \textbf{741}, 117-123 (2015)
doi:10.1016/j.physletb.2014.12.028
[arXiv:1410.6217 [hep-ph]].


\bibitem{arXiv6218}
A.~Celis, J.~Fuentes-Martín and H.~Serôdio,
JHEP \textbf{12}, 167 (2014)
doi:10.1007/JHEP12(2014)167
[arXiv:1410.6218 [hep-ph]].

\bibitem{geng11}  C-Q Geng and J. N. Ng, Phys. Rev. D41, 3848(1990).

\bibitem{sikivie}  P. Sikivie, Phys. Rev. Lett. 48, 1156(1982).

\bibitem{dw-problem}
Y. B. Zel'dovich, I, Y, Kobzarev, and L. B. Okun, Zh. Eksp. Teor. Fiz 67, 3(1974) [Sov. JETP 40, 1(1975)]; T. W. B. Kibble, J. Phys. A 9, 1387(1976).


\bibitem{georgi} H.~Georgi, D.~B.~Kaplan and L.~Randall,
Phys. Lett. B \textbf{169}, 73-78 (1986)
doi:10.1016/0370-2693(86)90688-X

\bibitem{new-axion} G.~Grilli di Cortona, E.~Hardy, J.~Pardo Vega and G.~Villadoro,
JHEP \textbf{01}, 034 (2016)
doi:10.1007/JHEP01(2016)034
[arXiv:1511.02867 [hep-ph]].



\bibitem{df} E.D. Commins and P.H. Bucksbaum, Weak interactions of leptons and quarks (Cambridge U.P., London, 1983).

\end{thebibliography}
\end{document}